\documentclass{webofc}
\usepackage[varg]{txfonts}   

\newcommand{\Mpl}{M_{\textrm{Pl}}}
\renewcommand{\(}{\left(}
\renewcommand{\)}{\right)}

\def\al{\alpha}
\def\bet{\beta}
\def\gam{\gamma}

\def\sig{\sigma}
\def\lam{\lambda}
\def\ep{\epsilon}

\def\N{\mathcal{N}}

\def\doi{http://doi.org}
\def\arxiv{http://arxiv.org/abs}
 \def\t{\tilde}
 \def\e{\mathrm{e}}
\def\r{\mathrm{r}}

\def\m{\mathrm{m}}

\def\s{\mathrm{s}}
\def\d{\mathrm{d}}

\begin{document}
\title{Quintessential inflation: A unified scenario of inflation and dark energy}
%
%

\author{\firstname{Md. Wali} \lastname{Hossain}\inst{1}\fnsep\thanks{\email{wali.hossain@apctp.org}} 
}

\institute{Asia Pacific Center for Theoretical Physics, Pohang 37673, Korea
          }

\abstract{%
Quintessential inflation unifies inflation and late time acceleration by a single scalar field. Such a scenario, with canonical and non-canonical scalar fields, has been discussed . The scalar field behaves as an inflaton field during inflation and as a quintessence field during late time. Also the predictions of the models has been compared with the recent Planck data.
}
\maketitle
%
\graphicspath{{figs/}}

\section{Introduction}
\label{intro}

Current observations suggest that there are two accelerated phases of our universe, one during very early time known as inflation \cite{Linde:1983gd} and the other one is the late time acceleration \cite{Copeland:2006wr}. These regimes of two accelerated phases are generally treated independently. However, it is interesting to think that there is a unique reason responsible for both phases of acceleration. Such a scenario is known as quintessential inflation \cite{Peebles:1999fz,Sahni:2001qp,Hossain:2014xha,Hossain:2014coa,Hossain:2014ova,Hossain:2014zma,Geng:2015fla}. 

The scalar field should not interfere with the thermal history of the universe. It should be invisible during radiation and matter dominated era and reappear only at the recent past giving rise to late-time cosmic acceleration. This requirement demands that the scalar field potential should have a steep region in the potential just after the end of inflation. This steep region results a scalar field kinetic energy dominated regime  which has two major effects, one is the fast decay in scalar field energy density which eventually results radiation dominated universe and the other one is the blue spectrum in relic gravitational wave \cite{Sahni:2001qp,Sahni:1990tx,Giovannini:1998bp}. Apart from these effects steep region of the potential can cause tracker behavior \cite{Steinhardt:1999nw} of the scalar field depending upon the fact that the scalar field potential has a shallow region during late time dynamics or there is a special mechanism, like non-minimal coupling, which helps the scalar field to dominate over matter. Since the scalar field must survive till late times the conventional reheating mechanism is not applicable and one needs alternative reheating mechanism. Instant preheating \cite{Felder:1998vq} mechanism is one of such efficient mechanisms.

In this article, I shall discuss quintessential inflation models with canonical and non-canonical scalar fields \cite{Hossain:2014xha,Hossain:2014coa,Hossain:2014zma,Geng:2015fla}. A non-minimal coupling between scalar field and massive neutrinos has been considered. This non-minimal coupling becomes effective only when the massive neutrinos become non-relativistic {\it i.e.}, only during the late times. in these scenarios the non-minimal coupling plays an important role in late time acceleration and the dark energy scale is related to the massive neutrino mass scale.

\section{Inflation}
\label{sec-1}
We shall first analyze the inflationary phase for both canonical and non-canonical scalar fields. We shall also compare the scenario with Planck data.

\subsection{Canonical scalar field}
\label{sec-can_inf}
Let us consider the following action \cite{Geng:2015fla}
\begin{eqnarray}
&&\mathcal{S} = \int d^4x
\sqrt{-g}\bigg[\frac{\Mpl^2}{2}R-\frac{1}{2}\partial^\mu\phi\partial_\mu \phi-V(\phi) \bigg]
\label{action0}
\end{eqnarray}
where $\Mpl$ is  the Planck mass,  $\phi$ is the scalar field and $V(\phi)$ is the scalar field potential. 

We consider the potential
 \begin{equation}
\label{potentialn}
V=V_0\e^{-\lambda \phi^n/\Mpl^n},
\end{equation}
where $V_0$, $\lam$ and $n$ are the constant parameters. $n>1$ makes the potential steeper than exponential which can affect the scalar field dynamics \cite{Geng:2015fla}. For $n>1$ the potential has a flat region around $\phi=0$ which can be responsible for the slow-roll of the scalar field. Increasing the values of $n$ increases the flat region in the potential which can give sufficiently low tensor to scalar ratio.  

Slow-roll parameters, for $n\neq 2$, are given below \cite{Geng:2015fla}
\begin{eqnarray}
\label{epsilon22}
\epsilon &=& \frac{\Mpl^2}{2}\(\frac{1}{V}\frac{\d V}{\d \phi}\)^2=\frac{1}{2}n^2\lambda^2\,Q(n,\lambda,\N)^{2n-2}\,, \\
\eta &=& \frac{\Mpl^2}{V}\frac{\d^2 V}{\d \phi^2} = n\lambda \,Q(n,\lambda,\N)^{n-2}\left\{1-n+n\lambda\, Q(n,\lambda,\N)^n \right\}\,, \\
Q(n,\lambda,\N)&=& \left\{n\lambda \Bigg((n-2) \N + n \lambda 2^{\frac{2 -n}{2 ( n-1)}}\left(\frac{1}{n^2 \lambda^2}\right)^{\frac{n}{2 ( n-1)}} \Bigg) \right\}^{1/(2-n)} \, .
\end{eqnarray}

\begin{figure*}
\centering
\includegraphics[width=6cm,clip]{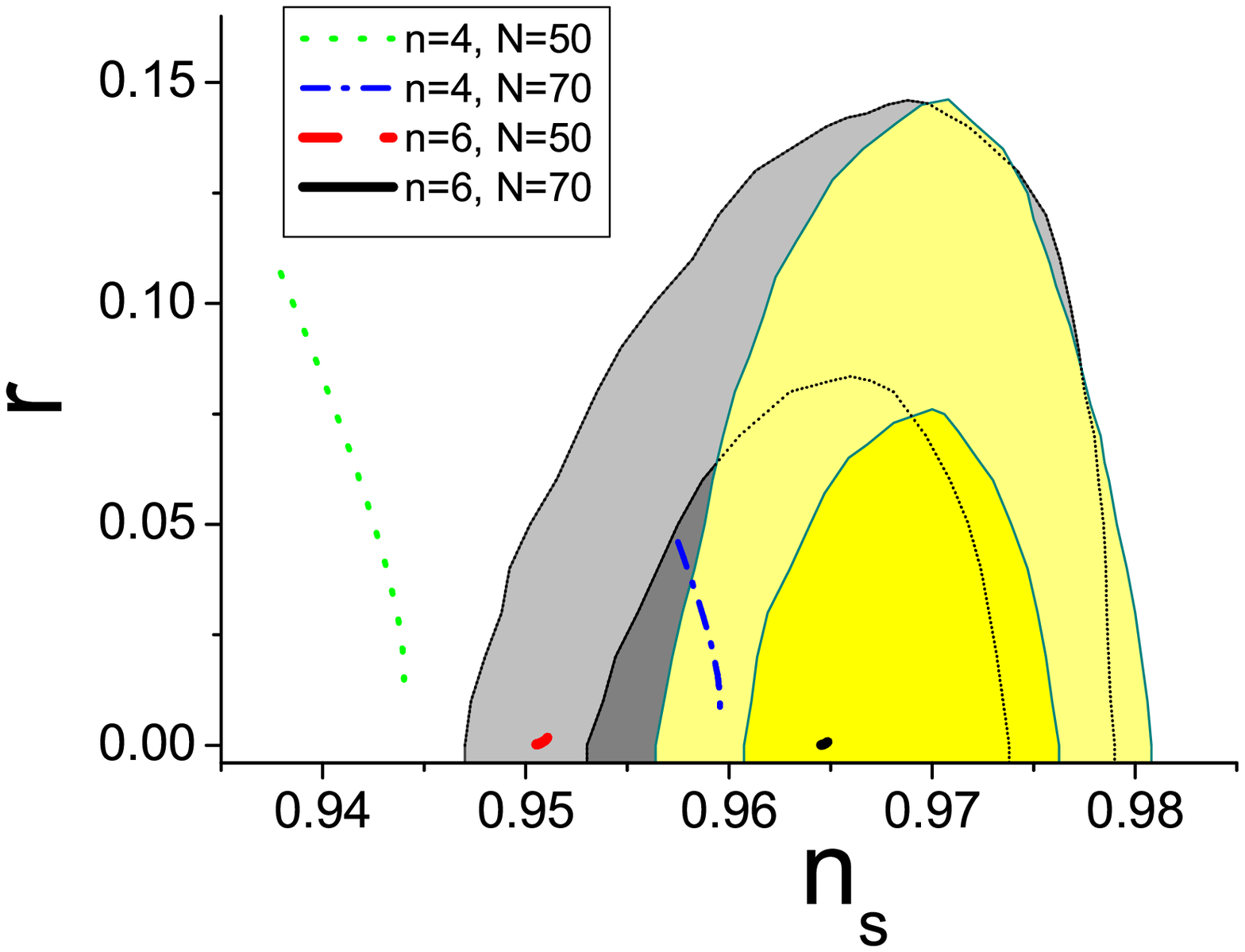}~~~~~~~~~
\includegraphics[width=6cm,clip]{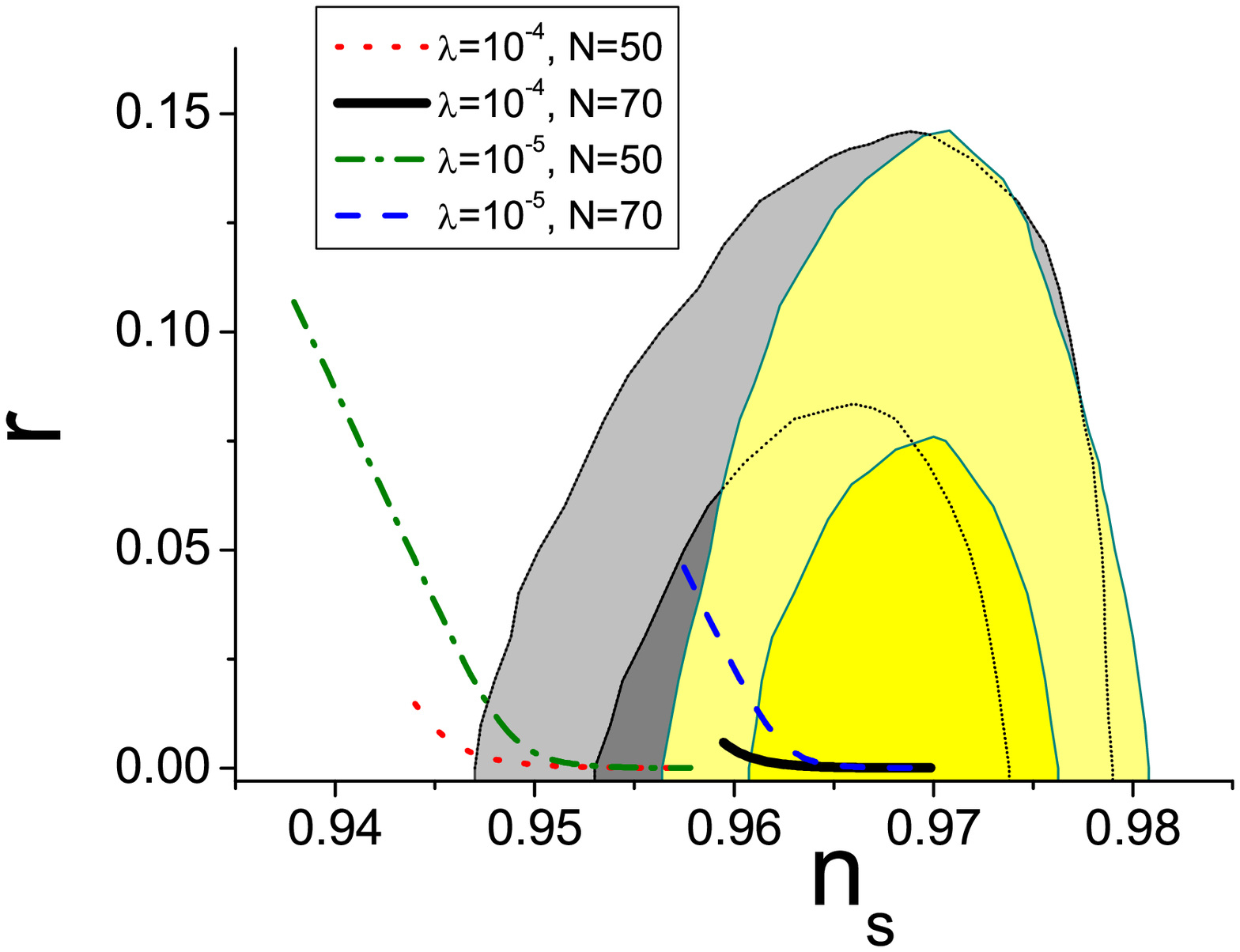}
\caption{1$\sig$ (yellow) and 2$\sig$ (light yellow) contours are for Planck 2015 results ($TT+lowP+lensing+BAO+JLA+H_0$) \cite{Ade:2015lrj} and gray contours are for Planck 2013 results  ($Planck+WP+BAO$) \cite{Planck:2013jfk}. Predictions of our scenario is shown by varying $\lambda$ for different $n$ (left figure) and varying $n$ for different $\lam$ (right figure) and for $50$ and $70$ e-foldings. Figures are taken from Ref.~\cite{Geng:2015fla}.}
\label{fig:can_inf}       
\end{figure*}

Tensor to scalar ratio ($r$) and spectral index ($n_\s$) are given by 
\begin{eqnarray}
&&
r(\mathcal{N})=16\epsilon(\mathcal{N})\, ,\\
&& n_\s(\N)= 1-6\ep(\N)+2\eta(\N) \, ,
\end{eqnarray}

In figure~\ref{fig:can_inf} the predictions of the model is being compared with the Planck 2015 \cite{Ade:2015lrj} and 2013 \cite{Planck:2013jfk} results. It can be seen that we can find some parameter space where the predictions of the model agree with the Planck results. 

\subsection{Non-canonical scalar field}
\label{sec-2}
We consider the following non-canonical action for the scalar field along with the standard Einstein-Hilbert term \cite{Wetterich:2013jsa,Hossain:2014xha},
\begin{eqnarray}
&&\mathcal{S} = \int d^4x \sqrt{-g}\bigg[-\frac{\Mpl^2}{2}R+ \frac{k^2(\phi)}{2}\partial^\mu\phi\partial_\mu \phi+V(\phi) \bigg],
~~~
\label{eq:action1}\\
&&k^2(\phi) = \(\frac{\al^2-\t\al^2}{\t\al^2}\)\frac{1}{1+\bet^2
\e^{\al\phi/\Mpl}}+1 \, , \\
&& V(\phi)=\Mpl^4\e^{-\al\phi/\Mpl} \, ,
\label{eq:pot_vg_phi}
\end{eqnarray}
where $\alpha$, $\tilde{\alpha}$ and $\beta$ are the model parameters. The kinetic function $k^2(\phi)$ has been chosen such a way so that we have tracker behavior in scalar field dynamics in post-inflationary periods. Non-canonical scalar field action can be transformed into a canonical form by doing a simple transformation $k^2(\phi)=(\partial \sig/\partial\phi)^2$ where $\sig$ is the canonical scalar field \cite{Hossain:2014xha}. In terms of the canonical field the potential will behave like exponential potential with slope $\t\al$ during inflation and $\al$ during post-inflationary dynamics. So by fixing $\t\al$ very small we can have a sufficiently flat region in the potential which can give rise to inflation with small tensor to scalar ratio. On the other hand $\al$ can be fixed from the constraint on the early dark energy. The parameter $\bet$ is related to the energy scale of inflation.

Tensor to scalar ratio ($r$) and spectral index ($n_\s$), for this scenario, are given by \cite{Hossain:2014xha,Hossain:2014coa,Geng:2015fla}
\begin{eqnarray}
&&
r(\mathcal{N},\t\al)=16\epsilon(\mathcal{N})\approx\frac{
8\t\al^2}{1-\e^{-\tilde\alpha^2\mathcal{N}}} \, ,
 \label{eq:r}\\
&& n_\s(\N,\t\al)= 1-6\ep+2\eta
\approx 1-\tilde{\alpha}^2\coth\(\frac{\t\al^2\N}{2}\)  \, ,
\label{eq:n_s}
\end{eqnarray}


\begin{figure}[h]
\centering
\sidecaption
\includegraphics[width=6cm,clip]{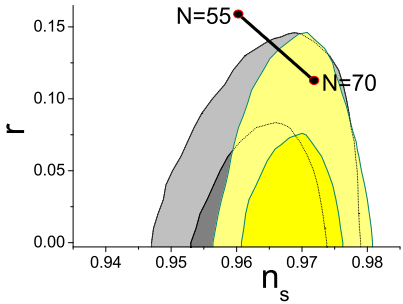}
\caption{1$\sig$ (yellow) and 2$\sig$ (light yellow) contours are for Planck 2015 results ($TT+lowP+lensing+BAO+JLA+H_0$) \cite{Ade:2015lrj} and gray contours are for Planck 2013 results  ($Planck+WP+BAO$) \cite{Planck:2013jfk}. Predictions of the scenario are shown by varying e-foldings from 55 to 70 for $\tilde{\alpha}\rightarrow0$. The figure is taken from Ref.~\cite{Geng:2015fla}.}
\label{fig:noncan_inf}       
\end{figure}

In figure~\ref{fig:noncan_inf} we have compared the model with recent observational results of Planck \cite{Ade:2015lrj,Planck:2013jfk} and found that the scenario under consideration can be ruled out by the future experiments as the model already lies in the $2\sig$ region of Planck 2015 results.

\section{Late time dynamics}
\label{sec:dark}

For both the cases, canonical and non-canonical, during the late time the potential is very steep. For canonical case the potential is steeper than exponential for $n>1$. For non-canonical case the potential is steep exponential as the function $k^2(\phi)\to 1$ for large values of $\phi$ {\it i.e.}, during the late time and the slope of the potential is steep as from the early dark energy constraints we get $\al\geq20$. Because of this steep nature of the potential the scalar field energy density follows the background during late time (scaling behavior) but can not take over matter as this behavior is an attractor solution for the potential with exponential nature. Here we should mention that the large field behavior of the potential~\eqref{potentialn} is similar as exponential potential \cite{Geng:2015fla} which gives same late time behavior of the scalar field as exponential potential. So for these scenarios under consideration we need some special mechanism so that the scalar field can take over matter at the recent past. In this regard we consider non-minimal coupling between scalar field and massive neutrinos. The motivation behind this is that the mass scale of massive neutrinos is of similar order of dark energy and in these scenarios we can relate the dark energy scale with the massive neutrinos energy density and the conformal coupling constant\cite{Hossain:2014xha,Hossain:2014zma,Geng:2015fla}. Another reason is that the massive neutrinos are relativistic during radiation and early matter dominated regimes and become non-relativistic only during late times. So the conformal coupling becomes effective only during the recent past as the scalar field is coupled with the trace of the energy momentum tensor of the massive neutrinos. We consider the conformal coupling of the form $\e^{2\gam\phi/\Mpl}$ where $\gam$ is the coupling constant. 

\begin{figure*}
\centering
\includegraphics[width=6cm,clip]{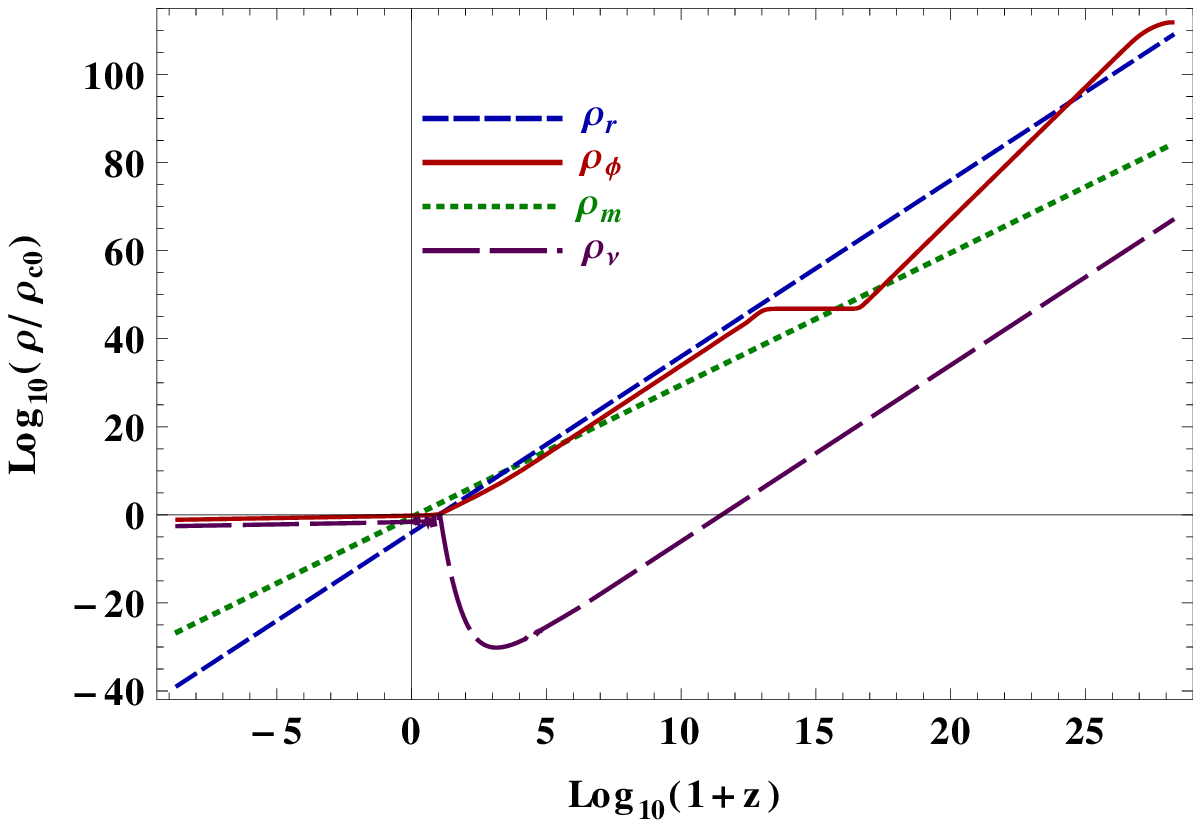}~~~~~~~~~
\includegraphics[width=6cm,clip]{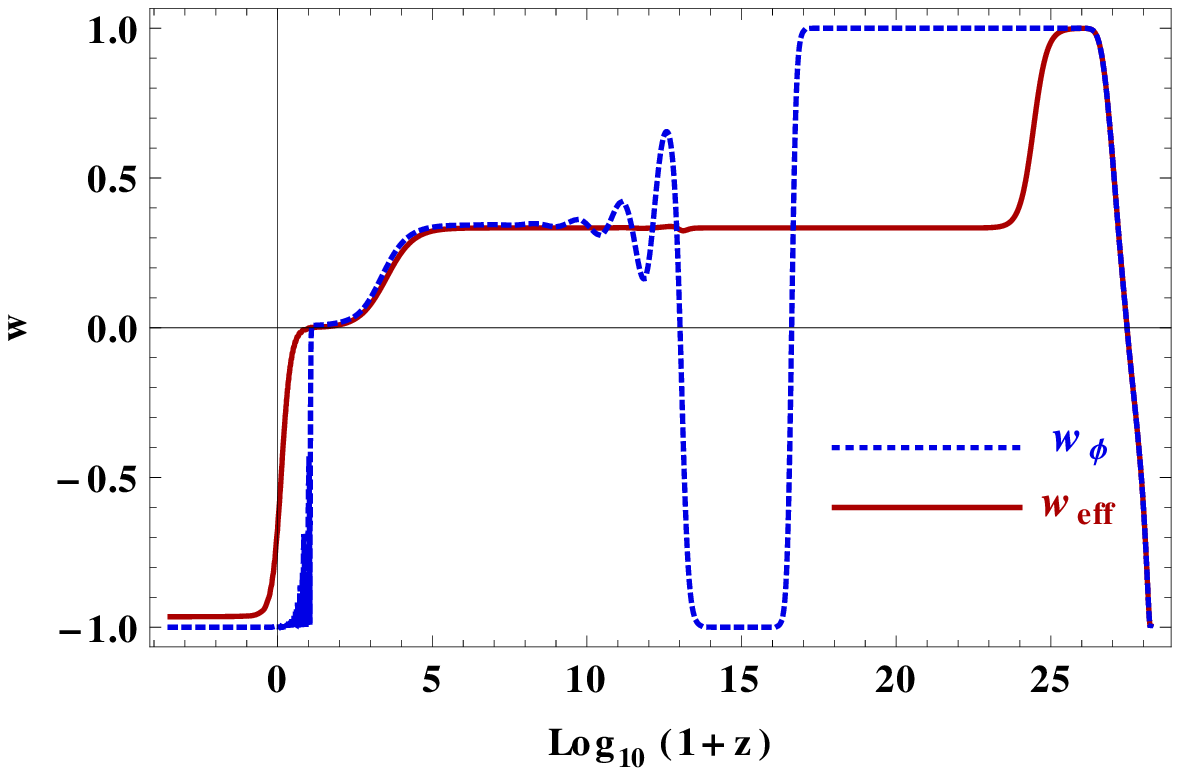}
\caption{Left: Evolution of matter ($\rho_\m$, dotted green), radiation ($\rho_\r$, blue short dashed), scalar field ($\rho_\phi$, red solid), and massive neutrinos ($\rho_\nu$, purple long dashed) energy densities. Right: Evolution of the effective ($w_{\rm eff}$, solid red) and scalar field ($w_\phi$, dotted blue) equation of states. Figures are taken from Ref.~\cite{Geng:2015fla}}
\label{fig:dark_energy}       
\end{figure*}

Non-minimal coupling modifies the scalar field equation of motion. This modification along with the potential term forms an effective potential. This effective potential has a minimum for $\gam>0$. This minimum depends on the massive neutrinos energy density and the conformal coupling constant. The scalar field, during late times, starts oscillating around the minimum. Now for large values of $\gam$ ({\it e.g.}, $\gam=800$) the minimum becomes very narrow and the scalar field eventually settles down at the minimum giving rise to late time acceleration. When the scalar field settles down at the minimum we can relate the dark energy density with the minimum of the effective potential and thereby with the massive neutrinos energy density and the conformal coupling constant. figure~\ref{fig:dark_energy} shows the evolution of the energy densities of different components of the universe during post-inflationary eras along with the evolution of the effective and scalar field equation of states. These figures are plotted by considering the canonical case. But it is needless to say that for the non-canonical case also we have similar dynamics of the scalar field because of the steep exponential nature of the potential.

\section{Summary and Discussions}
\label{sec:Summary}
Quintessential inflation models unify inflation and late time acceleration by a single scalar field. Models based on both canonical and non-canonical scalar fields are being discussed here. For canonical scalar field with potential steeper than exponential we can have low tensor to scalar ratio within the $1\sig$ bound given by Planck 2015 results. But for non-canonical scalar field the tensor to scalar ratio $>0.1$ which lies in the $2\sig$ contours of Planck 2015 results. In future, if we have more precise data, the non-canonical scalar field model of quintessential inflation, considered here, will be ruled out. 

Non-minimal coupling between the scalar field and massive neutrinos has been considered in both the scenarios. This non-minimal coupling plays an important role in late time acceleration. Since massive neutrinos are relativistic during early times and become non-relativistic only during the recent past the non-minimal coupling becomes effective only from the recent past. So the non-minimal coupling does not affect the radiation era. In both the scenarios, the late time behavior of the scalar field is a tracker behavior. Because of the non-minimal coupling there forms an effective potential with minimum and the scalar field oscillates around the minimum and eventually settles down at the minimum during late times giving rise to late time acceleration. 

Here we have presented models which can successfully unify inflation and late time acceleration by a single scalar field. While both of the scenario fit with the Planck 2015 results, at least up to $2\sig$ level, we expect that some of the models will be ruled out by future experiments.

\section*{Acknowledgment}
I thank C.~Q.~Geng, R.~Myrzakulov, M.~Sami and E.~N.~Saridakis for a fruitful collaboration on this topic. I also thank the organizers of the International Conference on Gravitation : Joint Conference of ICGAC-XIII and IK15, held at Ewha Womans University, Seoul, Korea, where the work was presented.


%
%
%

\end{document}